\documentclass[aps,prb,twocolumn,showpacs,showkeys,footinbib,amsmath,amssymb,superscriptaddress]{revtex4-1}

\usepackage{amsmath}
\usepackage{amssymb}
\usepackage{graphicx}
\usepackage{bm}
\usepackage{color}
\usepackage{relsize}
\usepackage{braket}
\usepackage{bbold}
\usepackage{txfonts}
\usepackage{epstopdf}
\usepackage{hyperref}

\makeatletter
    \renewcommand\@make@capt@title[2]{%
     \@ifx@empty\float@link{\@firstofone}{\expandafter\href\expandafter{\float@link}}%
      {\textbf{#1}}\@caption@fignum@sep#2\quad}%
    \makeatother
   
\makeatletter 
\renewcommand{\fnum@figure}{\textbf{Fig.~\thefigure}}
\makeatother

\begin{document}

\title{Fusion of Majorana bound states with mini-gate control in two-dimensional systems}

\author{Tong Zhou}
\email{tzhou8@buffalo.edu}
\affiliation{Department of Physics, University at Buffalo, State University of New York, Buffalo, New York 14260, USA}
\author{Matthieu C. Dartiailh}
\affiliation{Center for Quantum Phenomena, Department of Physics, New York University, New York 10003, USA}
\author{Kasra Sardashti}
\affiliation{Center for Quantum Phenomena, Department of Physics, New York University, New York 10003, USA}
\author{Jong E. Han}
\affiliation{Department of Physics, University at Buffalo, State University of New York, Buffalo, New York 14260, USA}
\author{Alex Matos-Abiague}
\affiliation{Department of Physics and Astronomy, Wayne State University, Detroit, Michigan 48201, USA}
\author{Javad Shabani}
\affiliation{Center for Quantum Phenomena, Department of Physics, New York University, New York 10003, USA}
\author{Igor \v{Z}uti\'c}
\email{zigor@buffalo.edu}
\affiliation{Department of Physics, University at Buffalo, State University of New York, Buffalo, New York 14260, USA}
\date{\today}

\begin{abstract}

A hallmark of topological superconductivity is the non-Abelian statistics of Majorana bound states (MBS), its chargeless zero-energy emergent quasiparticles. The resulting fractionalization of a single electron, stored nonlocally as a two spatially-separated 
MBS, provides a powerful platform for implementing fault-tolerant topological quantum computing. However, despite intensive efforts, experimental support for MBS remains indirect and does not probe their non-Abelian statistics. Here we propose how to overcome this obstacle in mini-gate controlled planar Josephson junctions (JJs) 
and demonstrate non-Abelian statistics through MBS fusion, detected by charge sensing using a quantum point contact, based on dynamical simulations. The feasibility of preparing, manipulating, and fusing MBS in two-dimensional (2D)
systems is supported in our experiments which demonstrate the gate control of topological transition and superconducting properties with five mini gates in InAs/Al-based JJs. 
While we focus on this well-established platform, where the topological superconductivity was already experimentally detected, our proposal to identify elusive non-Abelian statistics motivates also further MBS studies in other gate-controlled 2D systems.
\end{abstract}
\maketitle

\noindent{${\textbf{Introduction}}$}

\noindent{Proximity} effects can transform common materials to acquire exotic properties~\cite{Zutic2019:MT}.
A striking example is topological superconductivity hosting Majorana bound states (MBS)~\cite{Kitaev2001:PU,Fu2008:PRL,Lutchyn2010:PRL,Oreg2010:PRL,Klinovaja2012:PRL}.
Their non-Abelian statistics supports a peculiar state of matter, where quantum information stored nonlocally is preserved under local perturbation and disorder,  
particularly suitable for fault-tolerant quantum computing~\cite{Nayak2008:RMP,Aasen2016:PRX,DasSarma2015:NPJQI}.
Detecting MBS is mainly focused on one-dimensional (1D) systems~\cite{Mourik2012:S,Rokhinson2012:NP,Deng2012:NL,Nadj-Perge2014:S}
through spectral features, such as the zero-bias conductance peak (ZBCP)~\cite{Sengupta2001:PRB}.
However, even stable quantized ZBCP may not correspond to MBS~\cite{Zhang2021:X,Yu2021:NP,Pan2020:PRB}. 
While it is critical to identify MBS signatures that directly probe non-Abelian statistics,
1D systems require fine-tuned parameters for topological superconductivity~\cite{Lutchyn2010:PRL,Oreg2010:PRL} and  
limit probing non-Abelian statistics through MBS exchange (braiding) or fusion~\cite{Nayak2008:RMP,Aasen2016:PRX}.

Defects and quasiparticles in topological superconductors, or boundaries between topological and trivial regions, can bind localized Majorana zero-energy 
modes which behave as non-Abelian anyons~\cite{Aasen2016:PRX,DasSarma2015:NPJQI}. 
These zero-energy topologically-protected degenerate states, in which quantum information can be stored, are separated by the energy $\Delta$ from the excited states, 
as depicted in Fig.~1a. 
The ground states, nonlocally storing ordinary fermions, can be labelled by the fermion parity (even or odd), reflecting 0 or 1 fermion occupancy. 
For an ordinary fermion, $f$, composed of non-overlapping Majoranas, the ground state is twofold degenerate since both fermion parities correspond to zero energy. 
However, bringing the two Majoranas closer removes this degeneracy, as depicted in Fig.~1b and c. The resulting multiple fusion outcomes~\cite{Aasen2016:PRX,Lahtinen2017:SPP} 
\begin{equation} 
\gamma \times \gamma = I + \psi,
\label{eq:fuse}
\end{equation} 
reflect the underlying non-Abelian statistics and summarize that the fusion of the two MBS behaves either as vacuum,
$I$, or an unpaired fermion $\psi$, resulting in an extra charge. For the trivial fusion in Fig.~1b, when MBS with a defined parity within the same pair coalesce, the outcome is deterministic, it leads to the unchanged parity (shown to be even) with no extra charge.  For the nontrivial fusion in Fig.~1c both parities are equally likely,
a probabilistic measurement would yield an extra charge.  While a pioneering proposal for MBS fusion in 1D nanowires envisions gate-control
realization of Figs.~1b and c~\cite{Alicea2011:NP}, it has important obstacles.  
(i) Common nanowire geometries are surrounded by superconductors, the screening makes attempted gating ineffective. (ii) Topological superconductivity requires fine-tuned parameters~\cite{Lutchyn2010:PRL,Oreg2010:PRL}.  (iii) 1D geometry complicates detecting an extra charge from fusion. (iv) Without an accurate preparation of the initial state, the distinction between trivial and nontrivial fusion outcome is unclear. 

\begin{figure*}[t]
\centering
\includegraphics*[width=0.75\textwidth]{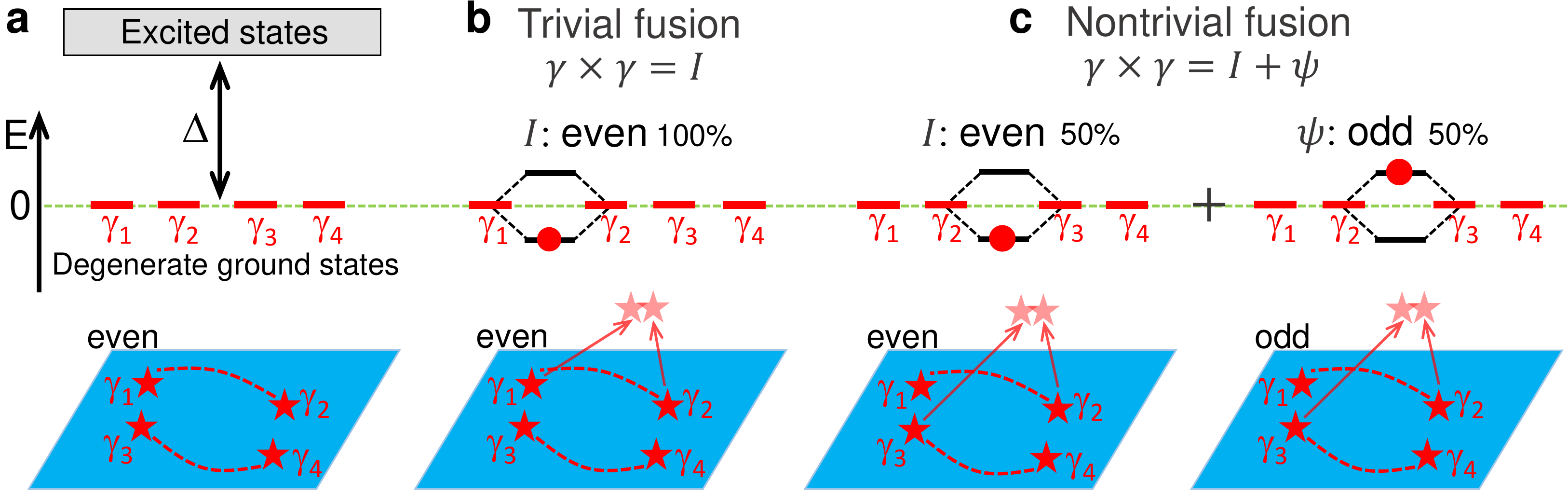}
\caption{\textbf{Schematic of the fusion rules.} {\bf a} Topological superconductor (blue) hosting 
MBS, $\gamma_{1,...,4}$. They behave as non-Abelian anyons and lead to the four-fold degeneracy in topological ground states, separated by the energy gap, $\Delta$, from the trivial excited states. {\bf b} and {\bf c} Different fusion outcomes: trivial fusion of $\gamma_1$ and $\gamma_2$, 100\% probability
to access vacuum, $I$ (Cooper pair condensate), and nontrivial fusion of $\gamma_2$ and $\gamma_3$, equal probabilities to access $I$ or an unpaired fermion,  $\psi$.  
Red dashed lines: paired MBS. In each case bringing closer MBS leads to the level splitting from the initial zero-energy modes.  
Filling the lower level, corresponding to $I$ with even parity, means the absence of a given particle, while filling the upper level
refers to $\psi$ with odd parity.
We assume initially even parity of the system. The net change in the charge characterizes the nontrivial fusion. 
}
\label{fig:VJF1}
\end{figure*}

\begin{figure}[t]
\includegraphics*[width=0.48\textwidth]{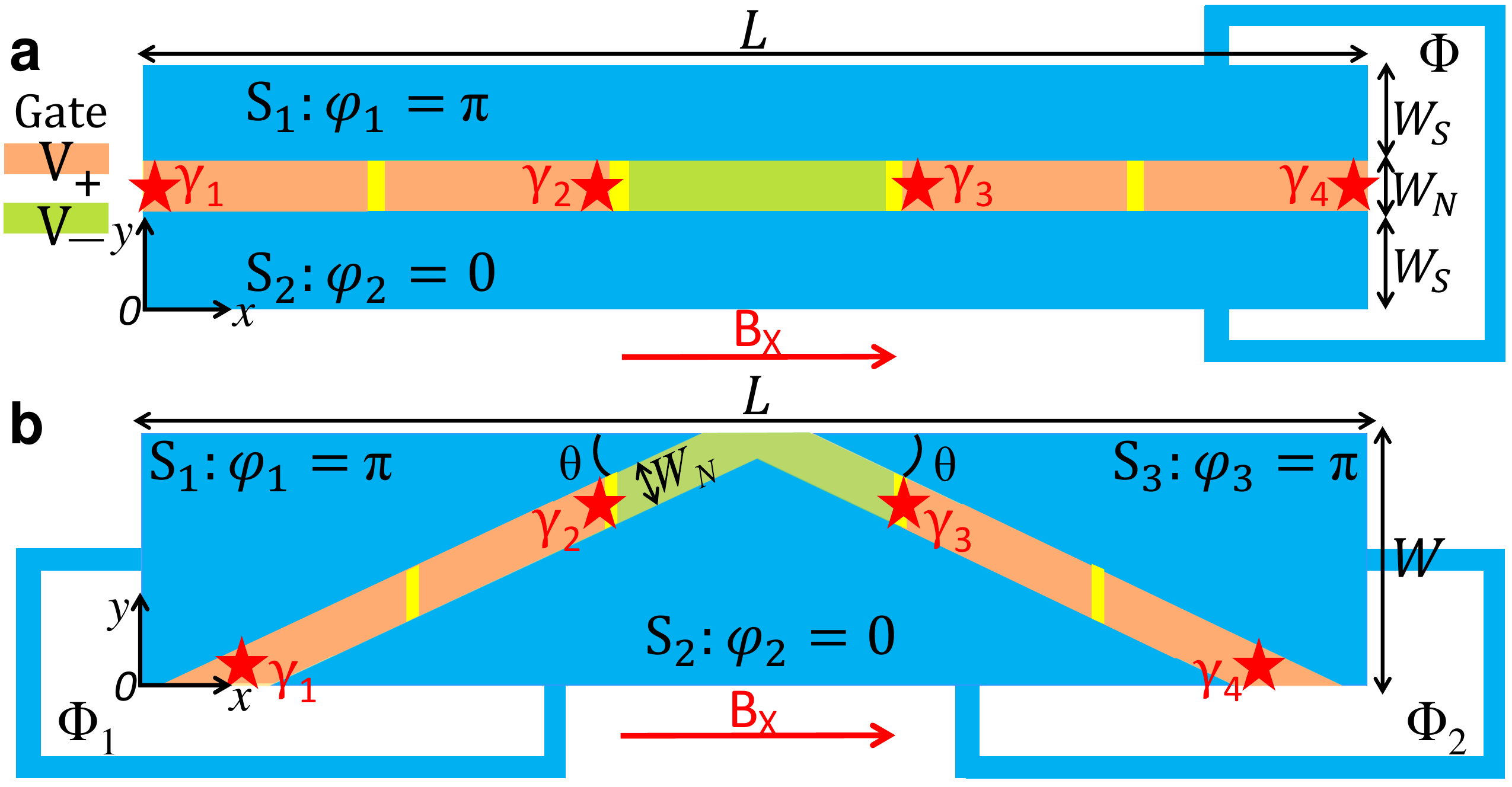}
\caption{\textbf{Setup.} {\bf a} Straight and {\bf b} V-shaped junction (SJ, VJ) formed by  superconducting, S$_{1,2,3}$, 
regions (blue), partially covering a 2D  
electron gas (yellow). The electron density in the uncovered part is locally tuned using mini-gate voltages, $V_{1,2,3,4,5}$
(left to right). With  a magnetic field B$_x$ and superconducting phases, $\varphi_{1,2,3}$, 
controlled by the external fluxes $\Phi$, $\Phi_1$, 
$\Phi_2$, the chemical potential of the normal region, $\mu_N$, is locally changed to support topological (orange) and trivial (green) regions, by imposing the  
mini-gate voltage $V_+$ and $V_-$. MBS $\gamma_{1,2,3,4}$ (stars) form at the ends of the topological regions.}
\label{fig:VJF2}
\end{figure}

Here we overcome these limitations by recognizing the opportunities in 2D proximitized 
materials~\cite{Shabani2016:PRB,Fatin2016:PRL,Matos-Abiague2017:SSC,Pientka2017:PRX,Hell2017:PRL,Scharf2019:PRB,Liu2019:PRB,Zhou2019:PRB,Setiawan2019:PRB2,Hegde2020:AN,Alidoust2021:PRB}. 
We reveal how mini-gate control in planar Josephson junctions (JJs) with 2D electron gas (2DEG) 
provides a versatile platform to realize MBS fusion. 
Our 2D InAs/Al JJs have proximitized 2DEG only partially-covered by superconductors. Mini-gates placed in the uncovered part strongly change the proximitized 
2DEG. Unlike fine-tuned parameters for 1D nanowires, recent experiments~\cite{Dartiailh2021:PRL,Fornieri2019:N,Ren2019:N} reveal that in planar JJs topological superconductivity exists over a large parameter space, and is particularly robust when the phase difference, $\phi$, between two superconducting regions is close to $\pi$. 

By proposing a V-shaped geometry, our JJ has its apex exposed edges where the locations of the bound states, formed through fusion, simplifies the charge detection in the adjacent quantum dot (QD) using quantum point contact (QPC)~\cite{Hanson2007:RMP, Barthel2009:PRL,Reilly2007:APL}. To distinguish the fusion outcomes in the charge detection, we reveal the importance of an accurate preparation of the initial state. We theoretically demonstrate the fundamental aspect of non-Abelian fusion that we can transform an MBS pair into an unpaired fermion, while using experimental parameters for topological superconductivity from our JJs~\cite{Dartiailh2021:PRL}. The feasibility of these findings is corroborated experimentally through the gate control of topological transition and superconducting properties and dynamical simulations of the MBS fusion.

Demonstrating fusion would be a major milestone for topological quantum computing and bridge the gap between the still
controversial MBS observation~\cite{Aguado2020:PT} 
and topological quantum algorithms,  largely detached from their materials implementation~\cite{Brown2017:PRX}. 
While the non-Abelian signatures from  MBS fusion are complementary to those obtained from 
braiding, experimentally the fusion is simpler.  There are even schemes in topological quantum computing 
implemented through fusion without 
braiding~\cite{Bonderson2009:AP,Litinski2017:PRB,Beenakker2020:SPP}.

\hspace*{\fill} \\
\noindent{${\textbf{Results}}$}

\noindent{${\textbf{Setup and model.}}$}

\noindent{Building} on our fabrications and experimental mini-gate control, we propose two geometries to fuse MBS,  
the straight and V-shaped planar Josephson junctions (SJ, VJ). 
Figure~2a   
shows the SJ setup,  
formed by two epitaxial superconducting layers covering a 2DEG with mini gates. A 1D normal region (N), defined between the superconducting leads $S_{1,2}$ with phases $\varphi_{1,2}$, can be tuned into the topological regime by the magnetic field B$_x$, the 2DEG chemical potential $\mu_N$ and the phase difference $\phi = \varphi_1 - \varphi_2$ between $S_{1,2}$, imposed by the magnetic flux $\Phi$. 
For $\phi\approx \pi$, the topological superconductivity exists over a large parameter space and is particularly robust~\cite{Hell2017:PRL,Pientka2017:PRX}. 
With $\phi=\pi$, for a certain B$_x$, the topological condition can then be directly controlled by the gate voltage through the changes in $\mu_N$~\cite{Hell2017:PRL}. We assume that gate voltage $V_+$ and $V_-$ support topological and trivial states, respectively. With mini gates, as depicted in Fig.~2a, we expect to electrostatically create multiple topological ($+$) and trivial ($-$) regions along the N channel by 
imposing the corresponding  voltage $V_+$ and $V_-$ in the five mini gates. Multiple MBS residing at the ends of topological regions can then be moved and fused. The setup of VJ is shown in Fig.~2b. 
It is similar to the SJ but has a V-shaped channel with an exposed apex defined by the three superconducting leads $S_{1,2,3}$. The corresponding phases $\varphi_{1,2,3}$ can be tuned by the magnetic flux $\Phi_{1,2}$. An advantage in the VJ is that its apex provides a place to detect the fusion outcome using QPC charge sensing.

Considering the topological condition for realistic planar JJs is complicated and strongly dependent on the system parameters~\cite{Dartiailh2021:PRL,Fornieri2019:N,Ren2019:N}, we need to explicitly calculate the relevant $V_+$ and $V_-$. To this end,
we simulate our fabricated planar JJs using the Bogoliubov-de Gennes (BdG) Hamiltonian,
\begin{equation}
\begin{aligned} 
H = \left[\frac{\mathbf{p}^2}{2m^\ast} - \mu_S + \mathbf{V}\left(x,y\right) + \frac{\alpha}{\hbar}\left(p_y\sigma_x - p_x\sigma_y \right)\right]\tau_z\\ - \frac{g \, 
\mu_B}{2}\mathbf{B}\cdot\boldsymbol{\sigma} + \Delta\left(x,y\right)\tau_+ + \Delta^\ast\left(x,y\right)\tau_-\;,
\end{aligned} 
\label{eq:BdG}
\end{equation}
where $\mathbf{p}$ is the momentum, $m^\ast$ is the effective electron mass, 
$\mu_S$ is the chemical potential in the considered $S_i$, $\alpha$ is the Rashba 
SOC strength, unless explicitly specified, 
$\text{B}\equiv \text{B}_x$. We use $\sigma_i$ ($\tau_i$) as the Pauli (Nambu) matrices in the spin (particle-hole) space and $\tau_\pm  = (\tau_x \pm i \tau_y)/2$. $\Delta(x, y)$  is the proximity-induced superconducting pair potential, for the 2DEG below the superconducting regions, 
which can be expressed, using the BCS relation for the B-field suppression, as
\begin{equation}
\Delta(x, y) = \Delta_{0} \sqrt{1-\left(\text{B} / \text{B}_c \right)^2}e^{i\varphi_i}, 
\label{eq:delta}
\end{equation} 
where  $\Delta_0$ is the superconducting gap at $\text{B}=0$, B$_c$ is the critical magnetic field, and $\varphi_i$ is the corresponding $S_i$  
phase.
The function $V(x, y) \equiv \mu_N(x, y) - \mu_S$ describes the local changes of $\mu_N(x, y)$  in the N region due to the application of the mini-gate voltages, $V_1,...,V_5$, as shown in Fig.~2.

\begin{figure}[t]
\centering
\includegraphics*[width=0.48\textwidth]{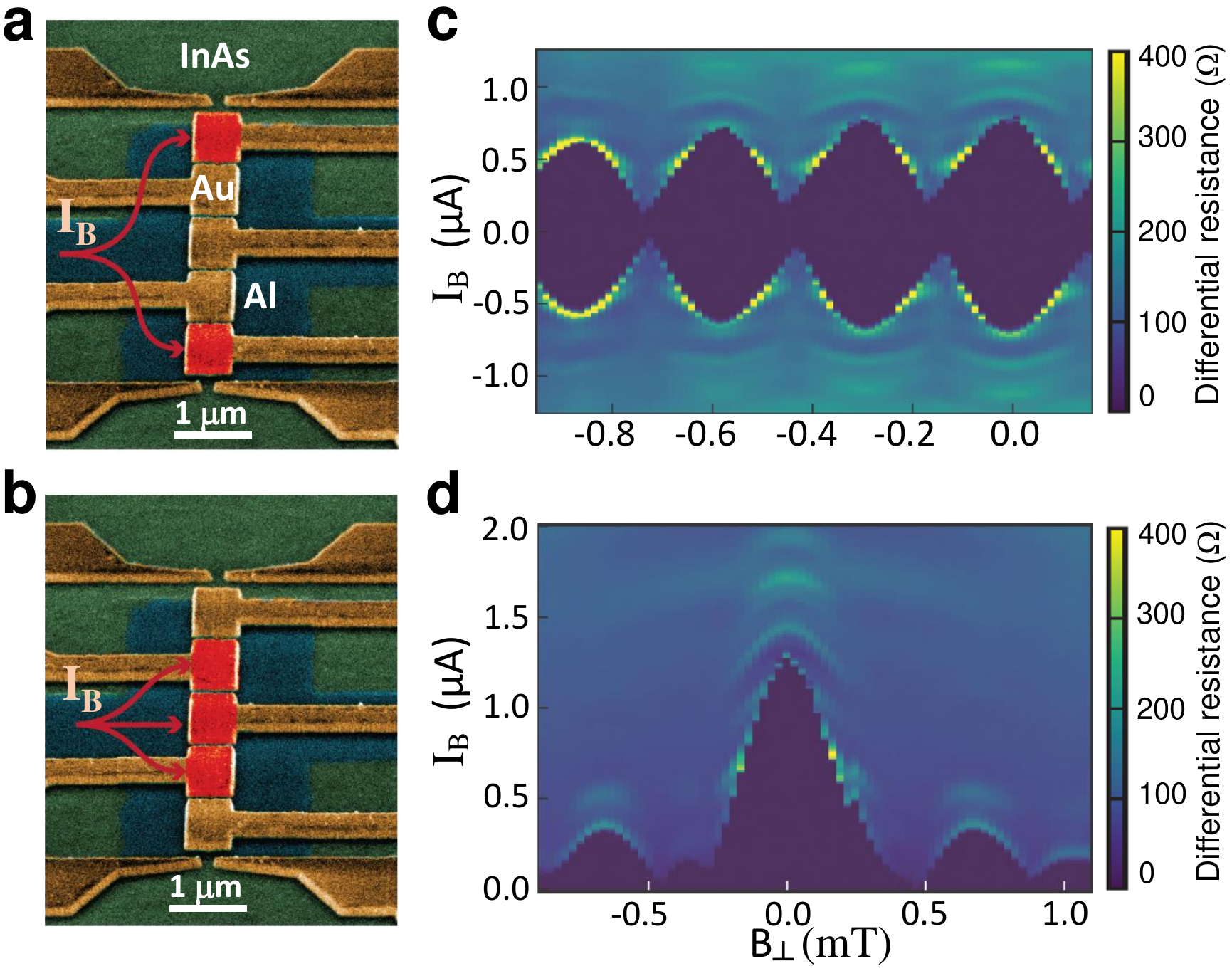}
\caption{\textbf{Experimental mini-gate control in an SJ.} {\bf a} and {\bf b} Scanning electron microscope (SEM)  image of an InAs/Al SJ with 5 mini gates covering the normal region. Mini gates can be controlled independently and those overlaid in red indicate the regions in which the applied bias current, I$_\text{B}$ can flow. The 2DEG is depleted under the other gates. {\bf c} and {\bf d} Differential resistance 
of the device as a function of the I$_\text{B}$ and out-of-plane B$_\perp$, corresponding to the gate configuration presented in {\bf a} and {\bf b}, respectively.}
\label{fig:VJF3}
\end{figure}

\begin{figure*}[t]
\centering
\includegraphics*[width=0.8\textwidth]{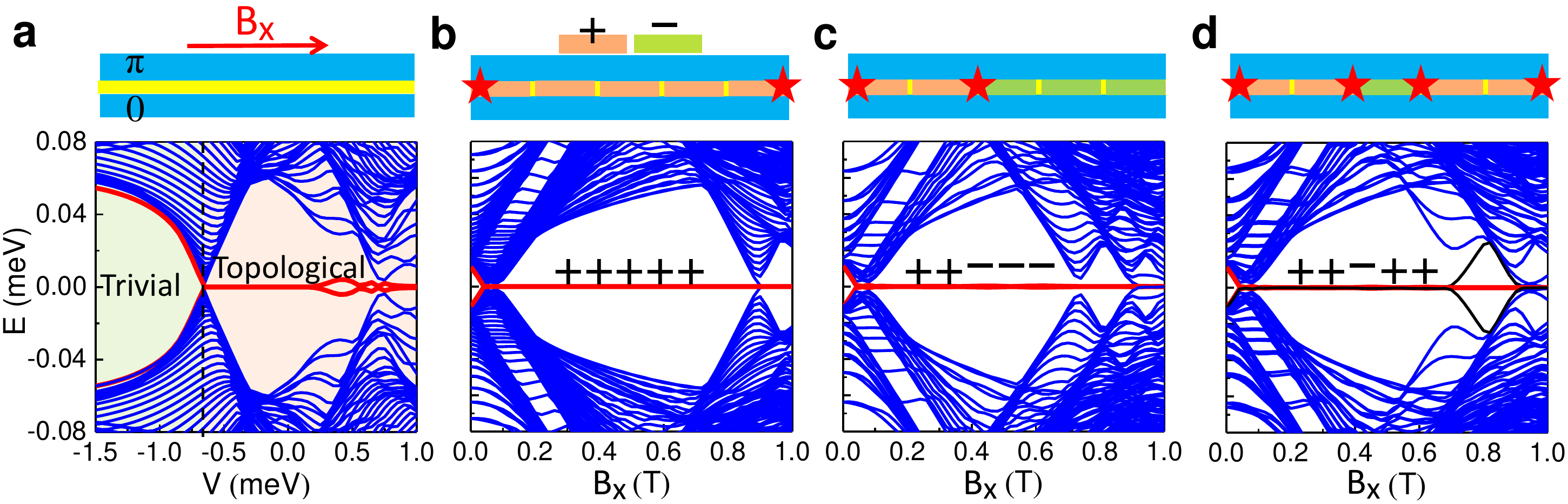}
\caption{\textbf{Mini-gate controlled topological transitions and MBS in an SJ.} {\bf a} Schematic (top) and energy spectra (bottom) for a planar $\pi$-SJ at B$_x = 0.6~$T as a function of the gate voltage ($V = \mu_N-\mu_S$), tuned by the top gate (not shown), which covers the whole N region. The black dashed line indicates the critical gate voltage between the trivial region (green) and topological region (orange). {\bf b}-{\bf d} Schematic (top) of a planar $\pi$-SJ with MBS (stars) for $+++++$, 
$++---$, and $++-++$ mini-gate configurations and the corresponding energy spectra (bottom), where $V_-= -1~$meV and $V_+=0~$meV are taken for $-$ and $+$ states, respectively. Red and black lines: evolution of finite-energy states into MBS inside the topological gap. The parameters are specified in the main text.}
\label{fig:VJF4}
\end{figure*}

In all the calculations, we choose the parameters consistent with our fabricated junctions (SJ and VJ) 
that also match experimental observation of robust proximity-induced superconductivity and topological states in epitaxial InAs/Al-based JJs~\cite{Dartiailh2021:PRL}, $m^\ast=0.03 m_0$, where $m_0$ is the electron mass, and $g=10$ for InAs, $\Delta_0$ = 0.23 meV, $\alpha = 10~$meVnm, B$_c$ = 1.6 T, and $\mu_S = 0.5~$meV. 
By switching 
$V_{+}$ and $V_{-}$ through mini-gate control, we expect to generate, manipulate and fuse MBS electrostatically. We will first demonstrate how this is realized in an SJ and then extend it
to a VJ to show how the QPC charge sensing can distinguish the trivial and nontrivial fusion.

\hspace*{\fill} \\

\noindent{${\textbf{MBS fusion in an SJ.}}$}

\noindent{An} experimental feasibility of the proposed mini-gate controlled MBS fusion builds on the demonstrated topological superconductivity in epitaxial InAs/Al planar 
JJs~\cite{Dartiailh2021:PRL,Barati2021:NL}. This is further corroborated by using the same platform to demonstrate that mini gates can modulate the superconducting
state in our fabricated SJ, shown with scanning electron microscope (SEM) images  in Figs.~3a and b.
With five gold mini gates covering the N region, $\mu_N$ for the each region under the mini gates can be independently tuned by the bias current, I$_\text{B}$.

With the three inner gates depleted, the current can only flow through the two outermost regions (marked in red) 
as depicted in Fig.~3a. In this configuration, the device behaves as a 
SQUID~\cite{Tinkham1996}, as seen from the map of the measured differential resistance as a function of I$_\text{B}$ 
and out-of-plane magnetic field in Fig.~3c which indicates interference between the current going though the two open channels. In contrast, when the three middle gates allow current to flow, and the outer most gates are used to deplete the 2DEG in Fig.~3b, the 
differential resistance in Fig.~3d shows a Fraunhofer pattern, typical of a single JJ~\cite{Tinkham1996}. As expected, its periodicity is close to the one of the SQUID configuration which contains the same region. 

Distinct features in  Figs.~3c and 3d 
show that locally $\mu_N$ 
is strongly changed by the mini gates, providing a clear advantage over an attempt of gate control in nanowire systems~\cite{Alicea2011:NP,Bauer2018:SP}, where the screening by superconductors diminishes changing $\mu_N$. 
Such gate-controlled superconducting response strongly supports  
our proposal of manipulating MBS with mini gates, when the topological superconductivity is achieved with B$_x$ and a phase bias, $\phi$. This demonstration of the mini-gate control, first established in our work, was later extended to experiments with even a larger number of mini gates~\cite{Elfeky2021:NL}.

Based on our fabricated device in Fig.~3, to obtain the relevant voltages $V_+$  ($V_-$) for topological (trivial) state, we do simulations based on the geometrical parameters depicted in 
Fig.~2a as $L=5~\mu$m, $W_S=0.3~\mu$m, $W_N=0.1~\mu$m, with each mini gate $1~\mu$m long. The calculated gate-voltage dependent energy spectrum with B$_x = 0.6~$T and $\phi = \pi$, is shown in Fig.~4a. The evolution of the lowest-energy states into zero-energy modes reveals that the 
MBS states emerge when the voltage exceeds the critical value $V_c = -0.7~$meV. 
This gives $V_+$ $\in$ ($-0.7~$meV, $1~$meV), confirmed by the spatially-localized 
probability density, $\mathrm{\rho_P}$, and the vanishing charge density, $\mathrm{\rho_C}$, while $V_- < V_c$ gives trivial states as shown in Supplementary Fig. 1. Such gate-controlled topological transition has been confirmed by the gap closing and reopening in our experiments as shown in Supplementary Fig. 2. We choose $V_+$ = 0 meV and $V_- = -1~$meV for the following simulations of mini-gate control. This identification of $V_+$ and $V_-$ gives us a chance to create 
and manipulate multiple MBS based on different mini-gate configurations. 

It is instructive to examine the topological robustness 
of the $+++++$ configuration, where all the mini gates are set at $V_+$, which is similar to a single topological SJ without mini gates. The whole N  
region is expected to be topological with MBS at its ends (Fig.~4b). The calculated B$_x$-dependent energy spectrum shows that MBS 
indeed exist in a very large range of B$_x$,
and a small B$_x\sim0.1~$T already supports MBS, in agreement with previous works~\cite{Hell2017:PRL,Pientka2017:PRX,Zhou2020:PRL}.
With mini-gate control  changing $+++++$ into $++---$, the MBS at the right end can be moved to the left part 
(Fig.~4c), while breaking the topological region into two separate ones, by changing $+++++$ into $++-++$,  
creates two MBS pairs  (Fig.~4d). These SJ configurations are revisited in Fig.~5, 
where we will see that the expected control of MBS is further corroborated by the calculated $\mathrm{\rho_P}$.

Following the above analysis, we propose a scenario for probing non-Abelian statistics based on fusion rules using mini-gate control as shown in the Fig.~5. 
The system is initially prepared in a trivial state (no MBS) with $-----$ configuration.  Subsequently, 
we can follow paths A and B to probe nontrivial and trivial fusion rules.  
For path A, in A$_1$ we first generate one MBS pair ($\gamma_1$, $\gamma_2$) by changing $V_1$ and $V_2$ from $V_-$ to $V_+$, and in A$_2$ the second MBS  pair 
($\gamma_3$, $\gamma_4$) by changing $V_4$ and $V_5$. These two MBS pairs build two complex fermions $f_{12}=\left(\gamma_1+\mathrm{i} \gamma_2\right) / 2$ and $f_{34}=\left(\gamma_{3}+\mathrm{i} \gamma_4\right) / 2$, which can be described by the occupation numbers $n_{12}$ and $n_{34}$. 
\begin{figure*}[t]
\centering
\includegraphics[width=0.8\textwidth]{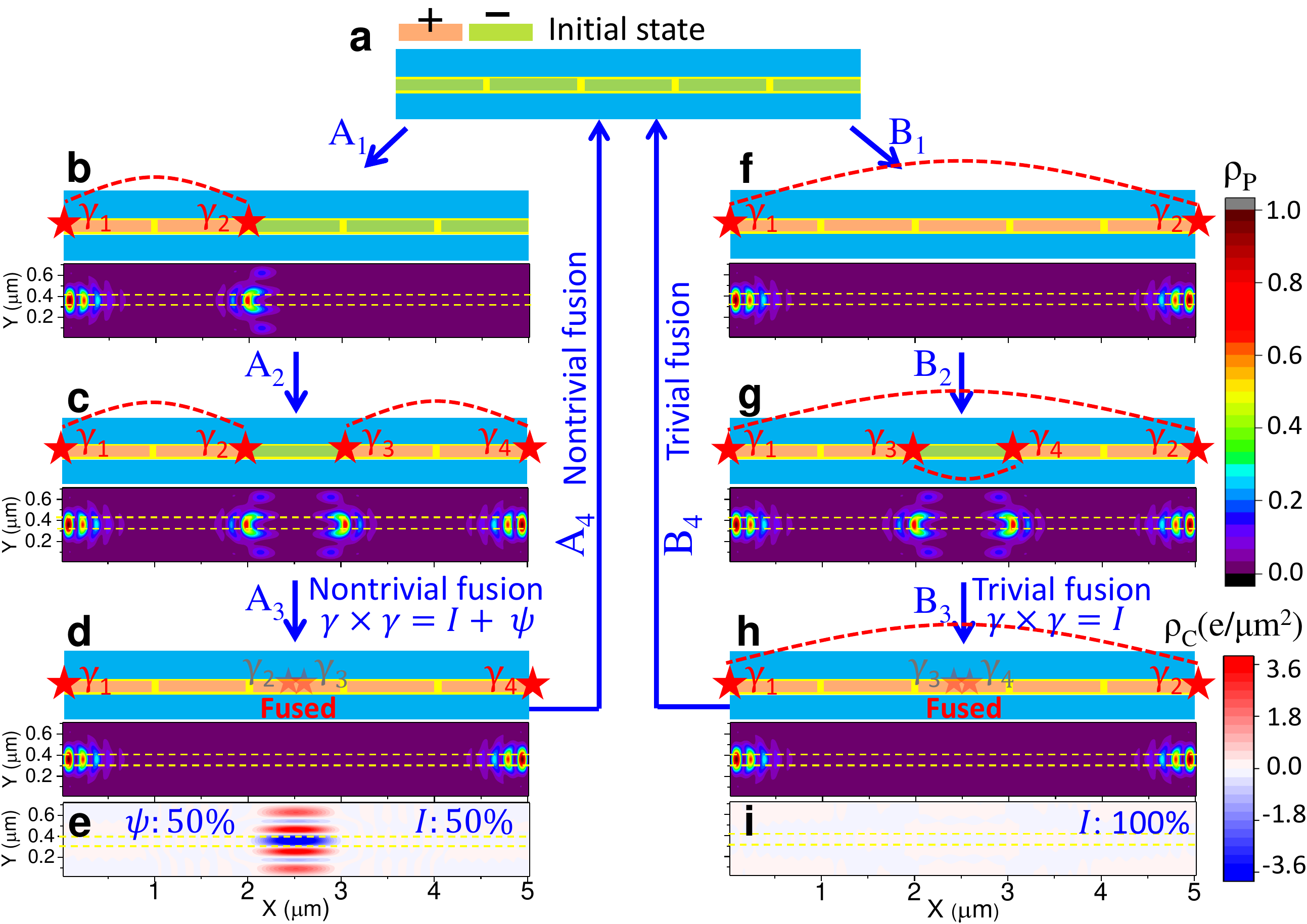}
\caption{\textbf{Probing non-Abelian statistics through MBS fusion in an SJ.} The scheme is supported by the calculated probability and charge densities, $\mathrm{\rho_P}$ and $\mathrm{\rho_C}$.
The red dashed lines link the same MBS pair, the yellow dashed lines indicate the N region covered by the mini gates.
{\bf a} Initial trivial state with $-----$ mini gates. {\bf b} A$_1$: changing $-----$ into $++---$,  MBS pair ($\gamma_1$, $\gamma_2$) is created. {\bf c} A$_2$: changing $++---$ into $++-++$, a second MBS pair ($\gamma_3$, $\gamma_4$) is created. {\bf d} A$_3$: changing $++-++$ into $+++++$, the MBS ($\gamma_2$, $\gamma_3$) are nontrivially fused at the center, accessing both 
vacuum, $I$, and an unpaired fermion, $\psi$, with 50$\%$ probability. For $I$, the system has no extra charge, supported by the vanishing 
$\mathrm{\rho_C}$ in {\bf i} for the ground state after the fusion.
For $\psi$ , the system has an extra charge, supported by the finite
sum of $\mathrm{\rho_C}$ in {\bf e} for the ground and first excited states after the fusion. {\bf f} B$_1$: changing $-----$ into $+++++$, the MBS ($\gamma_1$, $\gamma_2$) are created. {\bf g} B$_2$: changing $+++++$ into $++-++$, a second MBS pair 
($\gamma_3$, $\gamma_4$) is created. {\bf h} B$_3$: changing $++-++$ into $+++++$, the MBS ($\gamma_3$, $\gamma_4$) are trivially fused, corresponding to $I$ with 100$\%$ probability. A$_4$ or B$_4$: changing $+++++$ to $-----$, the remaining MBS pair is fused and the system returns to the initial mini-gate configuration. MBS fusion can be repeated following such operations. $\mathrm{\rho_P}$ is normalized to its maximum. The (minimum, maximum) values in {\bf e} and {\bf i} are (-3.5, 2.9) and (-0.00009, 0.00004), respectively.
All parameters are taken from Fig.~4.} 
\label{fig:VJF5}
\end{figure*}

\begin{figure*}
\centering
\includegraphics*[width=0.81\textwidth]{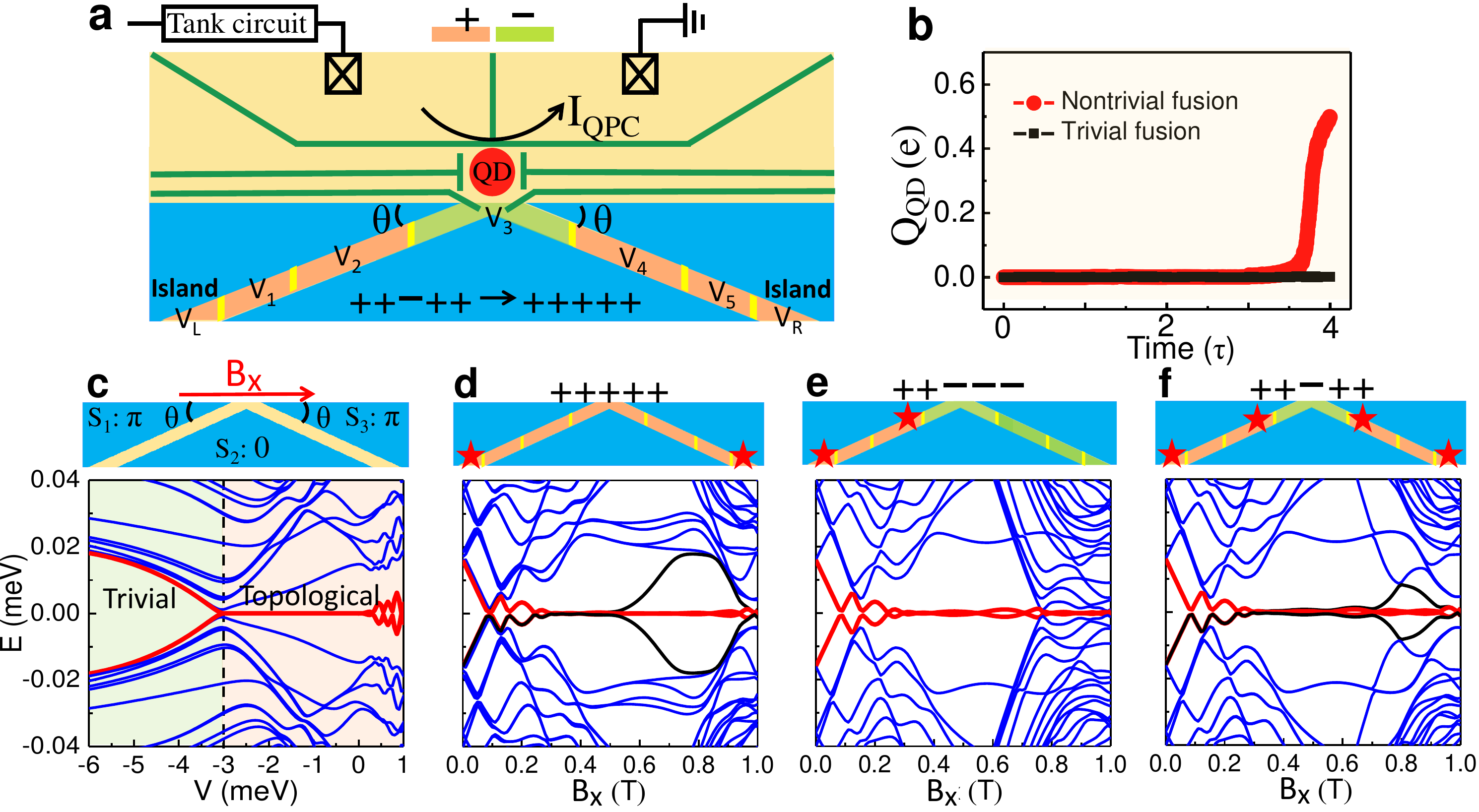}
\caption{\textbf{Mini-gate controlled topological transition and MBS fusion in the VJ.} {\bf a} Schematic of a setup with five mini gates and two quantum islands, $``L"$ and $``R"$, for the preparation of the initial states, with voltages $V_{1-5}$, $V_L$, and $V_R$. A quantum dot (QD) together with a quantum point contact (QPC) is coupled to the apex of the VJ to detect the fusion outcome. {\bf b} Calculated charge average of the QD, Q$_\text{QD}$, induced from the process of the trivial and nontrivial fusion, where $\tau$ is the mini-gates switching time with $1/\tau$ $\sim$ GHz. {\bf c} Schematic (top) and energy spectra (bottom) for a VJ with superconducting phases ($\pi$, 0, $\pi$) at B$_x= 0.7~$T as a function of the gate voltage $V$, tuned by the top gate (not shown), which covers the whole N region. The black dashed line indicates the $V_c$ between the trivial (green) and topological (orange) region. {\bf d}-{\bf f} Schematic (top) of a VJ with MBS (stars) for the $+++++$, $++---$, and $++-++$ configurations at B$_x = 0.7~$T and the corresponding energy spectra (bottom). Red and black lines: evolution of finite-energy states into MBS inside the topological gap. The geometric parameters are $L = 3.6~\mu$m, $W = 0.6~\mu$m, $W_N= 0.1~\mu$m, and $\theta = 0.1 \pi$. Other parameters are specified in the main text.}
\label{fig:VJF6}
\end{figure*}

Without loss of generality of demonstrating the fusion rules, we assume that the two fermion states are unoccupied, giving an initial state $\left|n_{12}, n_{34}\rangle=\right| 0_{12}, 0_{34}\rangle$. To facilitate experimentally probing the fusion rules, it is important to keep the same initial states in both trivial and nontrivial fusion. We will discuss later how to prepare the initial states by adding a new operation of $initialization$ before the MBS manipulations. In A$_3$ the change of $V_3$ from $V_-$ to $V_+$ nontrivially fuses ($\gamma_2$, $\gamma_3$), which accesses both the $I$ and $\psi$ fusion channels with equal probability. 
To better understand such nontrivial fusion, we reexpress the ground state in the basis of $f_{14}=\left(\gamma_1+\mathrm{i} \gamma_4\right) / 2$ and $f_{23}=\left(\gamma_2+\mathrm{i} \gamma_3\right) / 2$, i.e. $|\,0_{12}, 0_{34}\rangle=1 / \sqrt{2}\left(|\,0_{14}, 0_{23}\rangle-\mathrm{i} |\,1_{14}, 1_{23}\rangle\right)$, where 
$f_{14}f_{23} |\, 0_{14}, 0_{23}\rangle=0$, while $|\,1_{14}, 1_{23}\rangle=f_{14}^{\dagger} f_{23}^{\dagger}|\,0_{14}, 0_{23}\rangle$. Fusing ($\gamma_2$, $\gamma_3$) induces a finite energy to $f_{23}$, lifting the degeneracy between $|\, 0_{14}, 0_{23}\rangle$ and $|\,1_{14}, 1_{23}\rangle$. As a result, measuring such state then collapses the wave function with 50$\%$ probability onto either the ground state, $I$, or excited state with an extra quasiparticle, $\psi$. In A$_4$ fusing the remaining ($\gamma_1$, $\gamma_4$), by changing  
$+++++$ into  $-----$, drives the system  to the initial mini-gate configuration. 
To verify the non-Abelian statistics, we examine a trivial fusion scheme B$_1$-B$_4$. 
Unlike in the nontrivial fusion,  first ($\gamma_1$, $\gamma_2$) and then ($\gamma_3$, $\gamma_4$) are created by changing $-----$ to  $+++++$ and then to $++-++$. Therefore, fusing ($\gamma_3$, 
$\gamma_4$) can only access the $I$ channel with a trivial fusion because ($\gamma_3$, $\gamma_4$) belong to the same pair.

To simplify the description of MBS fusion it is helpful that considered scheme from Fig.~5 is adiabatic, which requires that the topological gap remains open during the
entire fusion. 
We show the corresponding evolution of the calculated low-energy spectra during the  
fusion in Supplementary Fig. 3. For any value of the continuously changing mini gates, the MBS are protected by the topological gap between the ground and first excited states which has the minimum value, $\Delta_\text{min} \approx 6~ \mu$eV. The $\Delta_\text{min}$ could be enhanced by increasing the Rashba SOC or using Sn or Nb with a higher bulk $\Delta$ than in Al ~\cite{Sau2010:PRB,Pakizer2020:P}. 
An animation for the evolution of the energy spectrum and wavefunction probability during the nontrivial fusion process is provided in Supplementary Movie 1.

Through uncertainty relations this $\Delta_\text{min}$ imposes a lower bound for the switching time, $\tau$, during the mini-gate operation, which can be estimated as $\tau_0 \sim \hbar/ \Delta_\text{min}$. In a realistic system, since the fusion involves multiple finite-size MBS pairs, their energies are not exactly zero and are characterized by their splitting, $\Delta_S$. Therefore, the switching time should be sufficiently short to ensure the non-adiabatic transition between these nearly-degenerate MBS levels, giving an upper bound $\tau < \tau_S = \hbar/ \Delta_S$. The upper bound is also constrained by the quasiparticle poisoning time, $\tau_\text{P}$. From the previous measurements in InAs/Al systems, $\tau_\text{P}$ was reported to be  between  $1~\mu$s and $10~$ms~\cite{Albrecht2017:PRL,Higginbotham2015:NP}. Together, $\tau_0 < \tau < \text{min}(\tau_S,\tau_P)$ is required for adiabatic fusion. In our SJ, this constraint implies $0.1~\text{ns} < \tau < 13~\text{ns}$, which is readily realized with the existing gate controlled employed in JJ-based qubits which are reaching GHz operation~\cite{Krantz2019:APR}.
The feasibility of this adiabatic evolution and distinct outcomes between the nontrivial and trivial MBS fusion are important prerequisites for using the fusion rules as an experimental verification of the non-Abelian statistics. A guidance for how the fusion rules could be measured comes from the prior proposals in nanowires, suggesting using Josephson current, fermion-parity, or cavity  detection~\cite{Aasen2016:PRX,Alicea2011:NP,Desjardins2019:NM,Rokhinson2012:NP,Ben-Shach2015:PRB,Wimmer2011:NJP}.
 
As shown in Fig.~1, the trivial fusion deterministically gives rise to the fusion channel $I$, preserving the charge of the system, while in the nontrivial fusion there is 50$\%$ probability for creating an extra charged quasiparticle $\psi$, which opens ways for charge
detection. We expect the dynamical process of the charge creation is associated with a Cooper pair which is then quickly absorbed into the spatially separated condensate for which the BCS formalism is adequate. Such an extra charge residing at a bound state [Supplementary Fig. 3] shows a huge local charge density difference compared to that in the $I$ fusion channel, which is verified by the four orders of magnitude difference in the corresponding $\mathrm{\rho_C}$ as shown in Figs.~5e and i. 
When the initial states are fixed, repeating operations A$_1$-A$_4$ from Fig.~5 is expected to give rise to charge fluctuations. In contrast, the fluctuations should be absent when repeating operations B$_1$-B$_4$ in the nontrivial fusion process. Detecting such charge fluctuations can be a direct evidence for the MBS nontrivial fusion and non-Abelian statistics.

\noindent{${\textbf{MBS fusion in a VJ.}}$}

\noindent{The} previous SJ geometry provides a plausible path to MBS fusion and distinguishing the resulting outcomes.
However, the corresponding charge fluctuations emerge in the interior of the central part of the N region, 
which is challenging to access experimentally due to the screening of superconductors and the presence of the top mini gates. Furthermore, it is unclear how to prepare the initial states, which is important to distinguish different experimental outcomes between the trivial and nontrivial fusion.

To overcome these difficulties, we propose a V-shaped geometry for the N-region where its apex is exposed to the edge, as shown in Fig.~2b. To control preparing initial states, we also add half-length mini gates ($``L"$, and $``R"$), with voltages  $V_L$ and $V_R$ at the ends of VJ, as shown in Fig. 6a. In their topological regime, $L$, and $R$, behave as effective quantum islands, supported by our calculations [Supplementary Note 1]. With two external fluxes, $\Phi_1$, $\Phi_2$, and mini-gate control, the MBS can be fused at the apex in a similar way to that in the SJ. An advantage in the VJ is that its apex provides a place to detect the additional charge induced by MBS fusion with QPC charge sensing, 
successfully used in semiconducting nanostructures~\cite{Hanson2007:RMP,Barthel2009:PRL,Reilly2007:APL}
 and also proposed for detection of topological superconductivity in 1D systems~\cite{Wimmer2011:NJP}. 
An experimental realization of the VJ with 5 mini gates, fabricated using standard electron-beam lithography and InAs/Al JJs, is shown in Supplementary Fig. 5.

A key difference from the SJ is that for the VJ, $B_x$ and the N/S interfaces are nor longer aligned. To support MBS in VJs, the topological superconductivity 
should survive to such a misalignment, characterized by the angle $\theta$ in Fig.~6a. As shown in Supplementary Fig. 4, our calculations reveal that topological superconductivity is supported for $\theta \leq 0.15\pi$. For a larger $\theta$, the topological states become eventually fully suppressed, consistent with the trends measured in planar JJs~\cite{Dartiailh2021:PRL}.
Based on the misalignment angle in the geometry of the fabricated VJ from Supplementary Fig. 5, 
we fix $\theta$ = 0.1$\pi$ in the following calculations.

The VJ geometry resembles a half of an X-junction~\cite{Zhou2020:PRL}, where various MBS can be created at the ends of the N regions by phase control.  
Similar as discussed for an SJ, a phase difference of $\pi$ between the two adjacent S regions supports topological superconductivity at a lower B$_x$. 
Therefore, as shown in  Fig.~6c, we fix the phases 
($\varphi_1$, $\varphi_2$, $\varphi_3$) of $S_1$, $S_2$, and $S_3$ as ($\pi$, 0, $\pi$) with external  fluxes $\Phi_1$ = $\Phi_2$ = 0.5$\Phi_{0}$, where $\Phi_{0}$ is the magnetic flux quantum, forming a $\pi$-VJ. A similar phase control with two external fluxes has been realized experimentally~\cite{Yang2019:PRB}.
Such a $\pi$-VJ is expected to exhibit topological superconductivity in the whole N region with MBS localized at its two ends. This can be seen in Fig.~6d when the gate voltage gives rise to topological states, analogous to the long-edge MBS in the X-junction~\cite{Zhou2020:PRL}. 

To identify the $V_+$ and $V_-$ in the $\pi$-VJ, we calculate the $V$-dependent energy spectrum at B$_x = 0.7~$T (see Fig.~6c). The evolution of the lowest-energy states show the critical $V_c$ = -3 meV in the VJ, where $V$ smaller (larger) than $V_c$ yields trivial (topological) states, further verified by the calculated $\mathrm{\rho_P}$ and $\mathrm{\rho_C}$ in Supplementary Fig. 6.
The chosen $V_+ = 0~$meV and $V_- = -5~$meV are used to manipulate the MBS with various mini-gate configurations. Similar to the SJ, for the $+++++$ configuration, the MBS are located at the ends of the N region, supported by the calculated zero-energy modes (Fig.~6d) and $\mathrm{\rho_P}$ (Fig.~7c). By changing $+++++$ into $++---$, the MBS can be moved to the left side (Fig.~6e), while changing $++---$ into $++-++$ creates another MBS pair on the right side (Fig.~6f). The zero-energy bands have small oscillations in the $++---$ and $++-++$ configurations because of the limited length of the topological regions.  These oscillations are suppressed by reducing the MBS overlap with an increased system size as in Supplementary Fig. 7.

\begin{figure*}
\centering
\includegraphics*[width=0.83\textwidth]{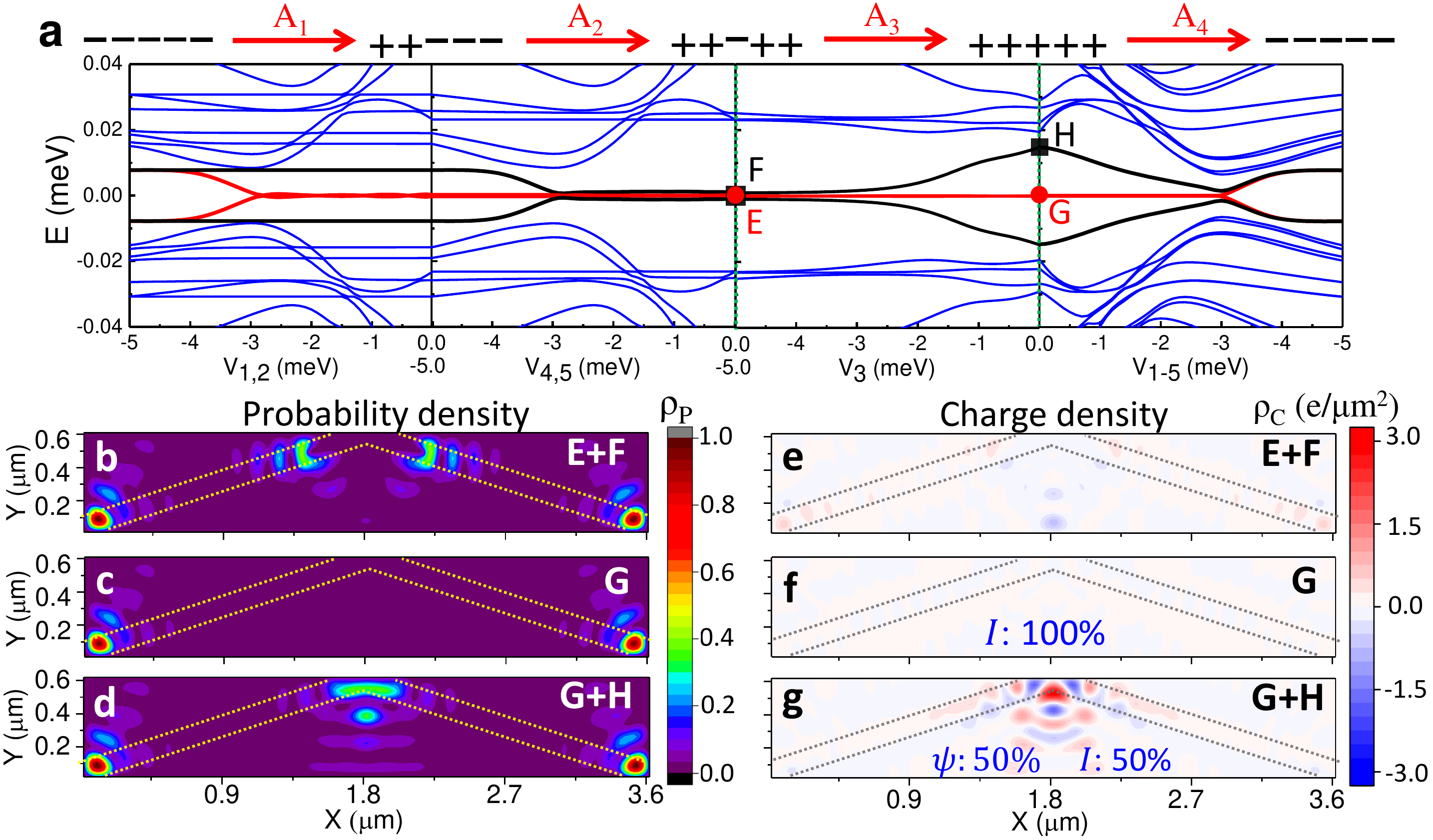}
\caption{\textbf{Outcome of the MBS fusion in a VJ.} {\bf a} Spectrum evolution for the operations A$_1$- A$_4$ as a function of the relevant mini-gate voltage. Red and black lines: evolution of finite-energy states into MBS inside the topological gap. 
E and F: the two MBS pairs (degenerate ground states) in the $++-++$ configuration (before fusion). G and H: the ground and first excited states in the $+++++$ configuration (after fusion).
{\bf b} Sum of the probability densities, $\mathrm{\rho_P}$ for E and F. {\bf c} $\mathrm{\rho_P}$ for G. {\bf d} sum of the $\mathrm{\rho_P}$ for G and H.
{\bf e}-{\bf g} The same as {\bf b}-{\bf d}, but 
for charge densities, $\mathrm{\rho_C}$.
The dashed lines mark the N regions covered by the mini gates. $\mathrm{\rho_P}$ is normalized to its maximum.
The (minimum, maximum) values in {\bf e}-{\bf g} are (-0.7, 0.5), (-0.02, 0.03) and (-1.9, 3.0), respectively. The parameters are taken from Fig.~6.} 
\label{fig:VJF7}
\end{figure*}

Similar to the fusion protocol in Fig.~5, the MBS trivial and nontrivial fusion can be implemented in a VJ as shown in Supplementary Note 2. The spectrum evolution during the nontrivial (trivial) fusion with A$_1$-A$_4$ (B$_1$-B$_4$) operations is shown in Fig. 7a (Supplementary Fig. 11). We can see the two quantum island states adiabatically evolve into two MBS pairs after the operations A$_{1-2}$/B$_{1-2}$. Such two MBS pairs, marked as E and F in Fig.~7a, are localized at the ends of the topological mini gates (Fig.~7b) in the $++-++$ configuration. 
They are chargeless before the fusion, supported by the calculated $\mathrm{\rho_C}$ in Fig.~7e. The operation $A_3$ fuses the MBS ($\gamma_2$ and $\gamma_3$) from different pairs and gives a nontrivial fusion. There is a 50$\%$ probability of attaining the ground state, $G$, localized at the ends of the N regions (Fig.~7c) with vanishing $\mathrm{\rho_C}$ (Fig.~7f), accessing the $I$ fusion channel. The other outcome, to attain with 50$\%$ probability the excited state, $H$, bound at the VJ apex (Fig.~7d), is accompanied with $\mathrm{\rho_C}$ (Fig.~7g) more than 
1000 times larger than that of the ground state at the VJ apex, accessing the $\psi$ fusion channel. In contrast, the operation $B_3$ trivially fuses the MBS ($\gamma_3$ and $\gamma_4$) from the same pair. The resulting outcome $I$ is achieved with 100$\%$ probability. 
Therefore, the probabilistic presence or absence of an
extra charge at the VJ apex is a signature of different fusion
outcomes.

To experimentally realize the fusions, the switching time should be tuned to enable that the MBS are adiabatically evolved during the whole fusion process in a VJ. The required switching time of the mini-gate control could be estimated analogously as for the SJ. We obtain $0.07~\text{ns} < \tau < 7~\text{ns}$ by calculating the  spectrum evolution during the whole fusion process (Fig. 7a), which is independently confirmed from our dynamical simulations shown in Supplementary Note 3.

The presence (absence) of the charge fluctuations when repeating nontrivial (trivial) is usually viewed as an evidence for the MBS fusion rules. However, if each time the initial state and its occupation changes randomly, the trivial fusion may also give charge fluctuations as a false signature of the fusion rules. This issue has been overlooked in previous fusion proposals~\cite{Aasen2016:PRX,Alicea2011:NP}, which neglect the effect of the initial occupations. To overcome this problem, as shown in Supplementary Note 1, we propose an $initialization$ A$_0$ (B$_0$), further supported by our simulations, to empty the initial occupations of the quantum island and get the $|00\rangle$ state. Such initialization precedes A$_1$-A$_4$ (B$_1$-B$_4$) operations to realize the initial $|00\rangle$ state in every fusion cycle, which corresponds to the operations A$_1$-A$_3$ (B$_1$-B$_3$) for the nontrivial (trivial) fusion. Observing the presence (absence) of the charge fluctuations at the VJ apex for repeated nontrivial (trivial) fusion can then be a conclusive evidence for the fusion rules.

\noindent{${\textbf{Readout of the fusion outcome.}}$}

\noindent{To} detect the charge fluctuations from the fusion, we couple a QD to the VJ apex. The QD is created by confining gates~\cite{Hanson2007:RMP}, as shown in Fig. 6a, and its energy levels can be shifted by the gate voltage $V_\text{QD}$. When the energy of the lowest unoccupied sate in the QD is aligned between the energy levels of the G and H states (Fig. 7a), the fusion-induced charge can transfer into the QD, giving a QD charge change, Q$_\text{QD}$. For trivial fusion, Q$_\text{QD}$ is 0; while for nontrivial fusion, Q$_\text{QD}$ is 0 or 1e with the same probability, giving an average value of 0.5e. Such a different fusion outcome is supported by the calculated Q$_\text{QD}$, shown in Fig. 6b, by using dynamical simulations based on time-dependent BdG equation~\cite{Amorim2015:PRB,Sanno2021:PRB}. More discussion and calculation details are given in Supplementary Note 3. The Q$_\text{QD}$ can be detected by the attached QPC~\cite{Hanson2007:RMP, Barthel2009:PRL,Reilly2007:APL}, because the QPC current, I$_\text{QPC}$, is very sensitive to the charge change~\cite{Barthel2009:PRL,Reilly2007:APL,Wimmer2011:NJP}. Such charge sensing technique has been widely used to accurately detect the charge in QDs~\cite{Hanson2007:RMP}. After the charge sensing detection, the fusion-induced charge does not stay in the VJ. We then reset the mini gates to the initial configuration (A$_4$/B$_4$) and do the initialization to make sure that each fusion cycle has the same initial state.

By repeating the operations of A$_0$-A$_4$ (B$_0$-B$_4$), we can repeat the MBS fusion with the same occupation state $|00\rangle$ as shown in the fusion protocols (Supplementary Note 2). Every time the fusion occurs (the system goes into $+++++$ after A$_3$/B$_3$), we use QPC charge sensing to detect $Q_\text{QD}$. The detected current, I$_\text{QPC}$, through the QPC is denoted by I$_\text{TF}$ (I$_\text{NF}$) for the trivial (nontrivial) fusion. While the expected I$_\text{TF}$ remains the same, the I$_\text{NF}$ fluctuates during the fusion cycles. To suppress the possible trivial background charge fluctuation, we can focus on the difference, $\Delta_\text{IF}$, between the I$_\text{TF}$ and I$_\text{NF}$.  Measuring such a fluctuating $\Delta_\text{IF}$ is a direct conclusive evidence for the non-Abelian statistics of MBS.

\hspace*{\fill} \\

\noindent{${\textbf{Discussion}}$}

\noindent{While} using the V-shaped junction requires some care in its design, such that magnitude of the misalignment angle between the N/S interface
and the applied in-plane magnetic field is not too large~\cite{Dartiailh2021:PRL,Zhou2020:PRL}, there are also important advantages of employing 
similar non-collinear structures to 
more completely manipulate MBS in 2D platforms and overcome the geometrical constraints of 1D systems. Within the same device 
footprint it is possible to pattern non-collinear structures where MBS are further separated and their hybridization is reduced to better attain the limit 
of chargeless zero-energy states. These 2D opportunities allow using zigzag structures for an improved robustness of MBS~\cite{Laeven2020:PRL} or creating 
multiple MBS~\cite{Zhou2020:PRL}. 
Progress in fabricating superconducting structures with topological insulators~\cite{Schuffelgen2019:NN,Yang2019:PRB} 
expands materials candidates to implement non-collinear JJs as 
platforms for MBS.

In the present work we have considered using the external flux control which can be conventionally realized through out-of-plane applied magnetic field. We have theoretically demonstrated the fundamental aspect of non-Abelian fusion that we can transform an MBS pair into an unpaired fermion, while using experimental parameters for topological superconductivity~\cite{Dartiailh2021:PRL}. Our experiments on mini-gate controlled superconducting properties in JJ and dynamical simulations of the MBS fusion are reassuring for the feasibility of these findings. However, future efforts may also take advantage of tunable magnetic textures as a method to implement a highly-localized flux control. Such textures could
be implemented with an array of magnetic elements or magnetic multilayers~\cite{Fatin2016:PRL,Zhou2019:PRB,Mohanta2019:PRA,Matos-Abiague2017:SSC,Palacio-Morales2019:SA,Turcotte2020:PRB,Ronetti2020:PRR,Wei2019:PRL}, 
as well as by using magnetic skyrmions~\cite{Yang2016:PRB,Gungordu2018:PRB,Garnier2019:CP,Mascot2020:npjQM}. 
The presence of magnetic
textures also extends the control of the spin-orbit coupling (SOC), beyond the usual classification into Rashba or Dresslhaus contribution~\cite{Scharf2019:PRB},
as such textures generate synthetic SOC~\cite{Fatin2016:PRL,Matos-Abiague2017:SSC,Kjaergaard2012:PRB}, and allow supporting MBS even in systems with inherently small SOC~\cite{Desjardins2019:NM,Turcotte2020:PRB}.

\hspace*{\fill} \\

\noindent{${\textbf{Methods}}$}

\noindent{${\textbf{Simulations.}}$}

\noindent{The} calculated results are obtained by numerically solving the BdG Hamiltonian from Eq. (2), using the Kwant package~\cite{Groth2014:NJP}. The dynamical simulations are performed by solving the time-dependent BdG equations~\cite{Amorim2015:PRB,Sanno2021:PRB}, as given in Supplementary Note 3.

\noindent{${\textbf{Fabrications.}}$}

\noindent{The} JJ structure is grown on semi-insulating InP (100) substrate, followed by a graded buffer layer. The quantum well consists of a 4 nm layer of InAs grown on a 6 nm layer of In$_{0.81}$Ga$_{0.25}$As. The InAs layer is capped by a 10 nm In$_{0.81}$Ga$_{0.25}$As layer to produce an optimal interface while maintaining high 2DEG mobility, followed by insitu growth of epitaxial Al (111). JJs are fabricated on the same wafer exhibit highly-transparent interface between the superconducting layer and the 2DEG. The fabrication process consists of three steps   of electron beam (e-beam) lithography using PMMA resist. After the first lithography, the deep semiconductor mesas are etched using first Transene type D to etch the Al and then a III-V wet etch [C$_6$H$_8$O$_7$(1M) : H$_3$PO$_4$(85$\%$ in mass) : H$_2$O$_2$(30$\%$ in mass) : H$_2$O = 18.3 : 0.43 : 1 : 73.3]. The second lithography is used to define the JJ gap which is etched using Transene type D. A layer of 90 nm of SiO$_x$ was then deposited using e-beam evaporation and finally the gates were patterned using e-beam lithography followed by e-beam evaporation of 5 nm of Ti followed by 45 nm of Au.

\noindent{${\textbf{Measurements.}}$}

\noindent{The} device has been measured in an Oxford Triton
dilution refrigerator fitted with a 6-3-1.5 T vector magnet which has a base temperature of 7 mK. All transport measurements are performed using standard dc and
lock-in techniques at low frequencies and excitation current I$_\text{ac}$  = 10 nA. 

\noindent{$\textbf{Data availability}$}

\noindent{The data that support the findings of this study are available within the paper and its Supplementary Information. Additional data are available from the corresponding authors upon reasonable request.}

\noindent{$\textbf{Code availability}$}

\noindent{The computation code information for getting the theoretical results is available from the corresponding authors upon reasonable request.}

\noindent{${\textbf{Acknowledgements}}$}

\noindent{We thank Jie Liu for helpful discussion. This work is supported by US ONR Grant No. N000141712793 (I.\v{Z}., J.H., and A.M.-A.), DARPA Grant No. DP18AP900007, and the University at Buffalo Center for Computational Research.}

\noindent{${\textbf{Author contributions}}$}

\noindent{T.Z. and I.\v{Z}. conceived the study. T.Z. performed the calculations and analysis with  J.H., A.M.-A., and I.\v{Z} providing input. M.D., K.S., and J.S. fabricated the samples and performed the experimental measurements and analysis. T.Z. and I.\v{Z}. wrote the paper. All authors were involved in discussion and editing of the paper.}

\noindent{${\textbf{Competing interests}}$}

\noindent{The authors declare no competing interests.}

\noindent{${\textbf{Additional information}}$}

\noindent{${\textbf{Supplementary information}}$} is available for this paper.

\noindent{${\textbf{Correspondence}}$} and requests for materials should be addressed to T.Z. (tzhou8@buffalo.edu) or I.\v{Z} (zigor@buffalo.edu).

\end{document}